\documentclass[aps,prb,preprint,showpacs,groupedaddress]{revtex4}
\usepackage[dvips]{graphicx}
\begin{document}
\title{Surface and interface study of pulsed-laser-deposited off-stoichiometric NiMnSb thin films on Si(100) substrate}
\author{S. Rai$^1$}
\author{M.K.Tiwari$^1$}
\author{G. S. Lodha$^1$}
\author{M. H. Modi$^1$}
\author{M. K. Chattopadhyay$^2$}
\author{S. Majumdar$^2$}
\author{S. Gardelis$^{3,6}$}
\author{Z. Viskadourakis$^{3,4}$}
\author{J. Giapintzakis$^{3,5}$}
\author{R. V. Nandedkar$^{1,2}$}
\author{S. B. Roy$^2$}
\author{P. Chaddah$^{2,7}$}
\affiliation{$^1$X-ray Optics Section, Synchrotron Utilization and Materials Research Division, Centre for Advanced Technology, Indore 452013, India.}
\affiliation{$^2$Magnetic and Superconducting Materials Section,\\ Synchrotron Utilization and Materials Research Division,Centre for Advanced Technology, Indore 452013, India.}
\affiliation{$^3$Foundation for Research and Technology-Hellas, IESL, 711 10 Heraklion, Crete, Greece.}
\affiliation{$^4$Department of Materials Science and Technology, University of Crete, Heraklion, Greece}
\affiliation{$^5$Department of Physics, University of Cyprus, 75 Kallipoleos, 1678 Nicosia, Cyprus.}
\affiliation{$^6$NCSR "Demokritos", IMEL, Athens 15310, Greece}
\affiliation{$^7$UGC-DAE Consortium for Scientific Research,\\ Khandwa Road,  Indore 452013, India.}
\date{\today}
\begin{abstract}
We report a detailed study of surface and interface properties of pulsed-laser deposited NiMnSb films on Si (100) substrate as a function of film thickness. As the thickness of films is reduced below 35 nm formation of a porous layer is observed. Porosity in this layer increases with decrease in NiMnSb film thickness.  These morphological changes of the ultra thin films are reflected in the interesting transport and magnetic properties of these films. On the other hand, there are no influences of compositional in-homogeneity and surface/interface roughness on the magnetic and transport properties of the  films. 
\end{abstract} 
\maketitle

\section{Introduction}
The half-Heusler compound NiMnSb is considered to be a half-metallic ferromagnet\cite{1} and a potential candidate as a spin-injector in spintronics devices\cite{2}. For this reason this system has been a subject of current interest and thin-films of NiMnSb have been successfully deposited on various semiconductor substrates\cite{3a,3b,4a,4b,4c,4d,4e}. While the question of spin-polarization across a metal-semiconductor interface still remains an open research problem, recent studies of electrical resistivity, magnetoresistance and Hall effect on off-stoichiometric NiMnSb films grown on Si substrate have revealed various interesting features \cite{5,6}. A low temperature upturn is observed in the temperature dependence of resistivity for film thickness 130 nm and below along with large positive magnetoresistance. As the film thickness decreases, the magnitude of both the resistivity upturn and the magnetoresistance increase\cite{5}. The low temperature resistivity upturn in the 5 nm sample is more dramatic than in the thicker samples, and the question arises whether another mechanism is operating in the thinnest sample\cite{5}.  In fact the nonlinearity in the I-V curve of the 5 nm film below the upturn is indicative of percolative behaviour in that film \cite{5}.  Hall effect measurements indicate that the room temperature electrical transport in these thin films becomes increasingly electron dominated with decreasing thickness\cite{6}; this is in marked contrast to the spin-polarized holes predicted for the bulk. Moreover the anomalous Hall conductivity cannot be interpreted within the usual scattering picture at low temperatures and this was attributed to a change in the spin dependent scattering\cite{6}. 

All these interesting features in transport properties raise the important question whether the nature of the films changes with the decrease in the film thickness and points toward the increasing significance of surface and interface of these films i.e. free surface electronic states.  This has motivated us for a thorough investigation of the surface and interface characteristics of these films.  Here we report a detailed investigation of surface and interface properties of these off-stoichiometric NiMnSb thin films grown on Si(100) substrate using grazing incidence x-ray reflectometry (XRR), grazing incidence x-ray fluorescence (GIXRF) spectrometry and energy dispersive X-ray fluorescence (EDXRF) spectrometry. Study of XRR provides information on the film thickness and electron density profile across the film. The depth dependent composition of these films is estimated using GIXRF and EDXRF spectrometry is used for the determination of bulk composition of the films. It has indeed been observed that the character of these films changes markedly below 35 nm thickness. Magnetization measurements have been done on some of these films to see the effect of thickness on the magnetic properties.  

\section{Experimental Details}
Four thin film samples of NiMnSb with nominal thickness 6, 34, 70 and 100 nm (estimation based on the number of laser pulses), were grown on Si(100) substrates using pulsed laser deposition (PLD)  at 475K from a slightly manganese- poor NiMn$_{0.95\pm 0.01}$Sb target. These samples will be referred to as A, B, C and D respectively in the discussion of the experimental results below. The details of sample preparations are given elsewhere\cite{3a,3b}.  The X-ray diffraction patterns taken with a thin film attachment in a Rigaku X-ray diffractometer are consistent with that of a NiMnSb half Heusler phase, and no second phase could be detected within the experimental resolution~\cite{5}. XRR measurements are carried out on a reflectometer developed in-house on a sealed tube with Cu target (= 0.154 nm). A near parallel incident beam was realized by using two slits of width 0.1 mm and 0.05 mm slit at the source and a razor blade was kept close to the sample surface to further reduce the beam size. Soller slit with 0.4$^{o}$ divergence were used in incident beam to control axial divergence. The reflected beam is analyzed using curved graphite monochromator and counted using scintillation counter. The final beam size in axial direction is 5 mm and the beam divergence was 0.025$^{o}$. The sample is mounted on a stage, which could be moved with 2.5-micron accuracy to bring the sample in the beam path. Before the measurements are carried out various standard alignment procedures are done to align both theta and 2$\theta$ axes within 0.005$^{o}$. All the measurements are performed with a step size of 0.01$^{o}$ in theta axes, which is sufficient to observe any small changes in thin films. Angle dependent GIXRF measurements are performed on a total external reflection X-ray fluorescence (TXRF) spectrometer developed in-house\cite{7}. The angle dependent fluorescence data are collected using a Peltier cooled solid-state detector. EDXRF measurements are done using a Cd$^{109}$ radioisotope excitation (22.4 keV Ag line) and a Si (Li) detector for x-ray fluorescence detection. Some complementary magnetization measurements have been performed using a commercial SQUID magnetometer (Quantum Design-MPMS5).

\section{Results and Discussion}

\subsection{X-ray reflectivity Measurements}
The XRR data are analyzed using the Parratt formalism\cite{8} to estimate the thickness and roughness of different layers of film. In the fitting procedure, one starts with a model structure consisting of layers with different thicknesses, roughness and densities. In this case if we start with a simple single layer model on Si substrate, the fit quality is not good. To improve the quality of fit we introduced more layers in a systematic manner and arrived at the best model, which yielded excellent fit with the measured reflectivity data. Figure 1 shows the best fit model extracted by fitting the reflectivity data. In this model, there are four layers comprising of 1.) native oxide layer on the silicon substrate, 2.) low density NiMnSb layer,  3.) NiMnSb layer with density near the bulk density and at the top, 4.) a low density porous layer (on the thinest films). The fit parameters, i.e. thickness, roughness and density of each layer, are allowed to vary in a controlled manner. The best fit gives the thickness, roughness and density of each layer. Using this information roughness convoluted scattering length density profile (SLDP) of the structure can be calculated.
 
\begin{figure}[t]
\centering
\includegraphics[width = 9 cm]{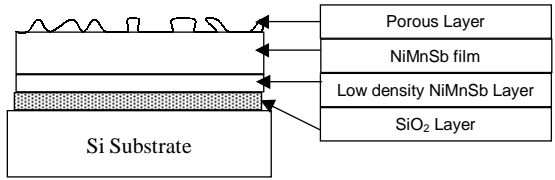}
\caption{The model used for fitting reflectivity data.}
\label{mt}
\end{figure} 
A non- linear least square-fitting algorithm was used to refine the thickness, roughness and density values by the $\chi^2$-minimization technique.  The best fits obtained are plotted in Fig.2 for each sample. Fig.3 shows roughness convoluted SLDP of the structure as a function of thickness for samples A to D obtained by fitting the measured reflectivity profile. 
\begin{figure}[t]
\centering
\includegraphics[width = 9 cm]{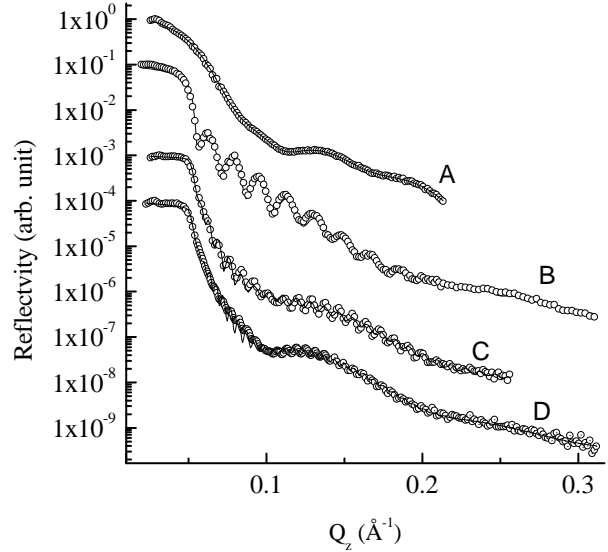}
\caption{X-ray reflectivity profile data (circular points) of samples A, B, C and D along with best fit line. The curves are vertically shifted for clarity.}
\label{rt}
\end{figure} 
The blank Si substrate roughness of 0.3 nm was independently estimated using XRR measurement. The details of each layer (thickness, roughness etc.) are given in table~\ref{layers} for samples A-D. Thickness of SiO$_{2}$ (native oxide) layer for all samples is 2-3 nm, which shows up as a dip in SLDP, marked as region 1 in Fig. 3. This layer is usually present due to oxidation of the top surface. Thickness of second layer i.e low-density NiMnSb layer is 3-5 nm for all samples, and density of this layer is 15-22 \% less than the bulk density, marked as region 2 in Fig 3. This low-density layer might have formed due to poor ordering of film leading to a less dense structure during initial growth process. This observation is in agreement with the observation of Schlomka {\it et. al.}\cite{9}. Above this low density layer there is NiMnSb film  with density  close to bulk density. Estimated thicknesses of NiMnSb film on samples A, B, C and D are 7.4, 33.2, 67.5 and 96 nm respectively. The error in thickness determination is less than 0.15 nm.  Note that these thickness values are not the same but close to those based on the number of pulses (nominal thickness values).  The roughness values are   within 1.2 to 1.8 nm.   For samples A and B, a reliable fit could not be obtained unless a porous layer (very low density layer) is incorporated in the model. This layer was not needed for samples C and D.  The SLDP for samples A and B shows (inset in Fig. 3) a fluctuating behavior at vacuum film interface. In sample A, the thickness of porous layer is 3 nm with roughness 0.8 nm . In the case of sample B roughness of the porous layer increases to 2.1 nm. which shows itself as a slow density gradient in SLDP.  The porosity of this layer for sample A is estimated to be 56 \% whereas for sample B it is reduced to 43 \%. The porous layer is not present in  samples C and D. This means that morphology of the film is changing with thickness. 
\begin{figure}
\centering
\includegraphics[width = 9 cm]{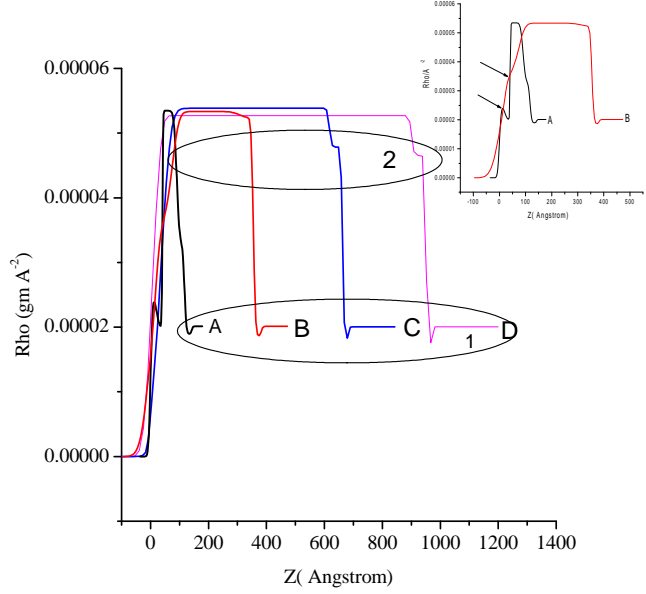}
\caption{Scattering length density profile (SLDP) of structure as a function of thickness for samples A, B, C, and D. Inset SLDP of sample A and B is shown magnified for clarity.}
\label{lnr}
\end{figure} 

\subsection{EDXRF and GIXRF  Measurements}
For the determination of trace impurities all of the NiMnSb samples are analyzed using EDXRF spectrometry. As the excitation energy is 22.4 KeV, there is no K excitation of Sb and the L excitation is very weak; actually the L fluorescence line of Sb could not be resolved in our case.  The EDXRF measurements show that there are no other trace impurities present in the samples. Ni to Mn mass concentration ratio in all the samples is found to be 1.17$\pm$0.18. 

\begin{table}
\centering

\begin{tabular}{|p{2.5 cm}|p{3cm}|p{4.1cm}|p{3cm}|p{4cm}|} 
\hline
 {\bf Sample Name}& {\bf SiO$_2$ Layer}   & {\bf Low density NiMnSb layer}  & {\bf NiMnSb Layer}& {\bf Porous Layer}  \\ 

 & (Thickness, Roughness)& (Thickness, Density,Roughness) & (Thickness, Roughness)& (Thickness, Density, Roughness) \\ \hline \hline
 
Sample A  & 3.8 nm, 0.7 nm & 2.5 nm, 22\% less, 1.0 nm &4.9 nm, 1.2 nm& 2.9nm, 56 \% less, 0.8 nm   \\  \hline

Sample B & 3.0 nm 0.5 nm & 4.5nm, 15 \% less, 1.9 nm & 28.7 nm, 1.6nm & 3.0 nm, 43\% less, 2.1 nm \\ \hline

Sample C& 2.2 nm, 0.4 nm & 4.4 nm,15 \% less, 0.5 nm & 63 nm, 1 nm& Not Present \\ \hline

Sample D& 2.1nm, 0.4 nm & 5.0 nm, 15\% less , 0.8 nm & 90 nm, 1.3 nm & Not Present \\ \hline

\end{tabular}
\caption{Details of thickness, roughness and density of various layers of all  NiMnSb/Si(100) films (Samples A-D). Densities of {\it Low density NiMnSb} and {\it porous  layers} are gives as percent change of {\it NiMnSb layer}.}
\label{layers}
\end{table}
The thickness and the density of NiMnSb are further estimated for sample B by measuring the angle dependent fluorescence. With Cu K$_\alpha$ excitation energy we observe Sb L$_{\alpha}$ and Mn K$_{\alpha}$ fluorescence, but no K excitation from Ni. In the present measurement, angle dependent GIXRF intensity measurements are model fitted using a rectangular concentration depth profile for the element of interest. The model is based on a matrix formalism, which accounts for standing wave fields due to Fresnel transmission and reflection amplitudes in layered material. Fig. 4 shows the variation GIXRF intensities for Sb-L$_{\alpha}$ and Mn-K$_{\alpha}$ in sample B as a function of incident angle, along with the best fit obtained as a result of iterative model calculations. The estimated film thickness is $\approx$ 31 nm and the density is close to bulk density of NiMnSb. The difference in thickness value is due to constant density model used for fitting GIXRF data. The maximum fluorescence intensity is at an incidence angle 0.39$^o$ for both elements. This indicates that the composition of Mn and Sb is constant across the depth of the film. In case of segregation of Ni, Mn or Sb in the film we would have observed two different critical angles. These GIXRF results substantiate that density fluctuations observed by x-ray reflectivity profile fitting are due to structural discontinuities such as porosities and not due to density variations due to segregation of Ni, Mn or Sb atoms in the film. 
\begin{figure}[t]
\centering
\includegraphics[width = 9 cm]{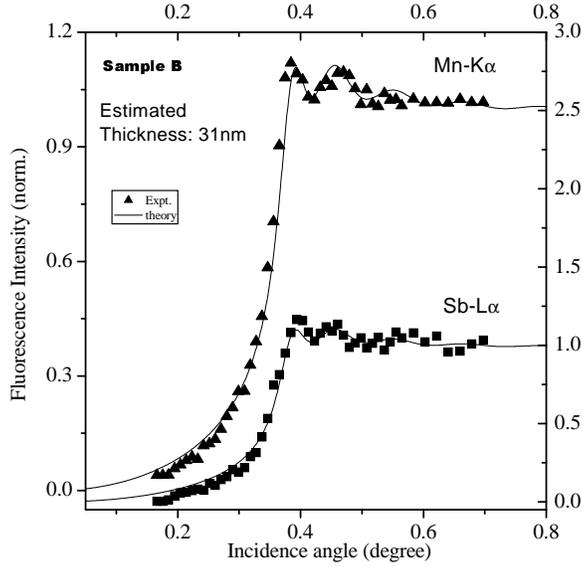}
\caption{GIXRF profile measured for a sample B at incident excitation energy E$_0$ = 8.047 keV. The scattered points represents measured data where as solid line shows model fitted GIXRF profile. The normalization was done at higher angle ($\sim$ 1.0 degree).}
\label{rth}
\end{figure} 
     
\subsection{Resistivity and magnetization measurements}
To check whether this difference in morphology / density profile cause any influence on the physical properties of these thin films, resistivity and magnetization measurements have been performed on the thickest sample D and thinnest sample A. As in the earlier study\cite{5} the resistivity measurements show a distinct difference between these two samples. The thinnest sample A shows a negative coefficient of resistance in the temperature regime down to 30 K, clearly indicating non metallic character of the sample. While indication exists for non-metallic behavior in the sample D also in the low T regime, at higher $T$ ($>$ 200 K) it definitely shows metallic behavior. A detailed magneto-transport measurement on these thin films is in progress (Ref. Results to be published).

In magnetization measurements, the magnetic contribution from the blank Si substrate is measured first in the temperature region 5-300 K in the presence of various applied magnetic fields. It shows a diamagnetic response in the entire temperature range. We then measure the magnetic response of the complete sample unit (thin film+substrate) under the same experimental condition and subtract out the contribution from the Si-substrate to get the actual magnetic response of the NiMnSb thin films. The magnetization ($M$) versus field ($H$) plot (See Fig.5) of sample D at 5K clearly shows that it is ferromagnetic in nature. The finite slope of $M-H$ curve above the technical saturation point ($H$=2kOe) is indicative of the disordered nature of the ferromagnetic state. 
\begin{figure}[t]
\centering
\includegraphics[width = 9 cm]{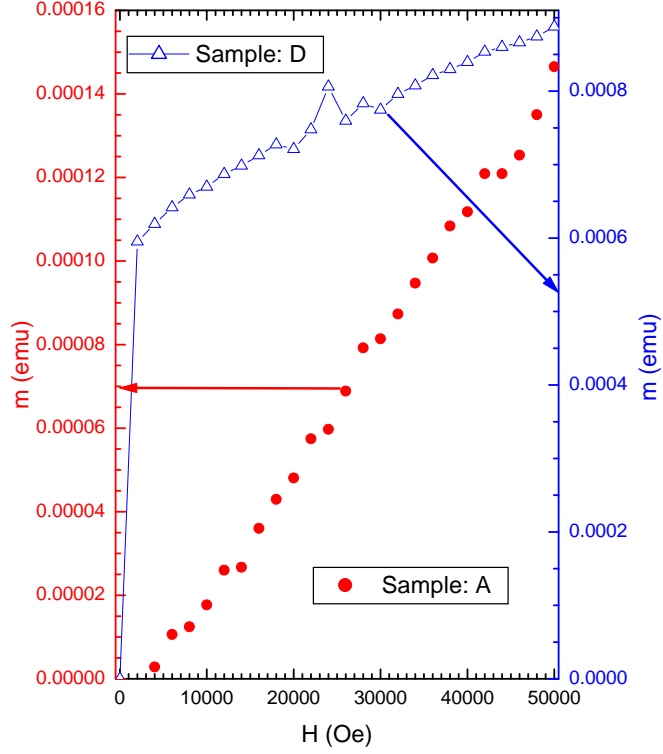}
\caption{Magnetization versus field plots for NiMnSb/Si samples A and D at 5K}
\end{figure}
There is a hint of a peak in the temperature($T$) dependence of $M$ in this sample below 10 K (see Fig.6). However, $M$ decreases continuously between 10 and 300K and the general nature of the $M$ versus $T$ plot (Fig.6 ) is commensurate with the ferromagnetic state of the sample. In contrast the $M-T$ curve for the thinnest sample A shows a  peak around 80 K. Further, the almost linear $M-H$ curve at 5 K (see Fig.5) clearly indicates that this sample is not ferromagnetic in nature. 
\begin{figure}[t]
\centering
\includegraphics[width = 9 cm]{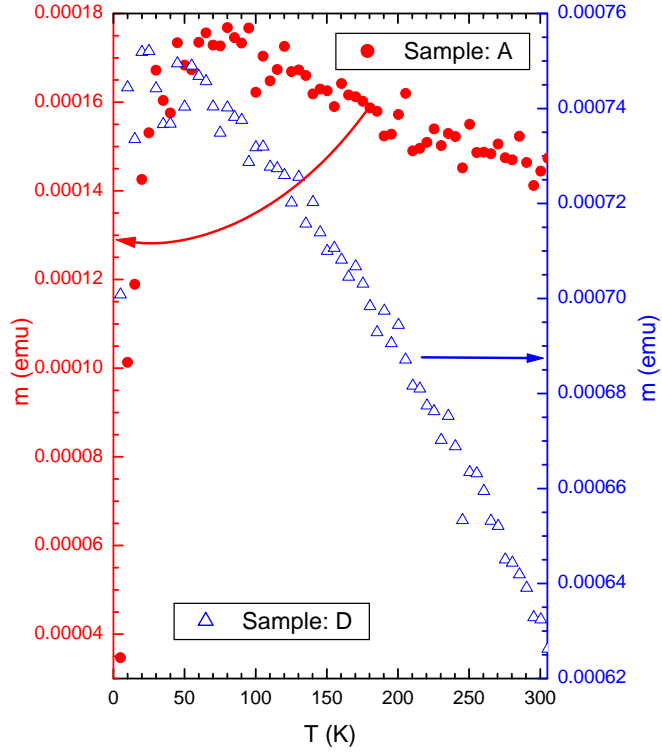}
\caption{Magnetization versus temperature plots for NiMnSb/Si samples A and D with 50 kOe of applied magnetic field.}

\end{figure}
The long range ferromagnetic order of the film-D is further established through $M^2$ versus $H/M$  Arrott plot presented in fig.7. The positive intercept of this plot clearly indicates the presence of spontaneous magnetization in this sample. While more experiments are necessary to ascertain the exact nature of the magnetic state of the sample A, it can definitely be said that the ferromagnetic state of bulk NiMnSb which is still prevailing in the thicker film D, does not exist in this sample.  These results strongly suggest that the morphology indeed influences the physical properties of these films. In the light of these results, the earlier suggestion of percolative behavior in the transport properties of the ultra thin films of NiMnSb on Si substrate\cite{5} gains much strength.  
\begin{figure}
\centering
\includegraphics[width = 7 cm]{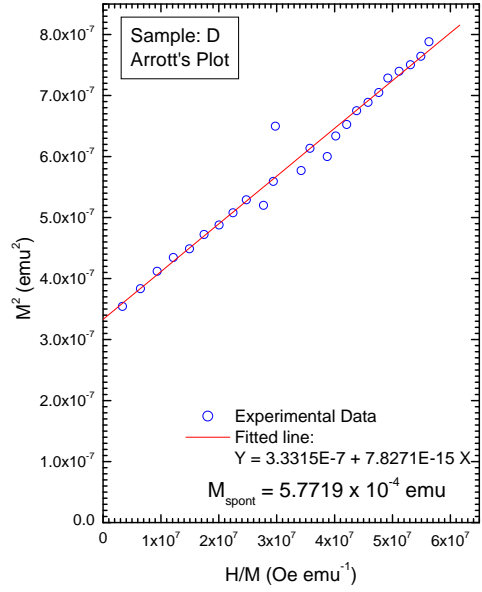}
\caption{Arrot plot for NiMnSb/Si sample D at 5K}
\label{mr}
\end{figure}

\section{Conclusions}
XRR studies on off-stoichiometric NiMnSb thin film samples grown on Si substrate show the formation of a very low-density layer at the top for the 7.6 and 33.2 nm thick films. GIXRF measurements clearly indicate the absence of any segregation of alloying elements. Hence this layer must be a highly porous layer, which is appearing as a very low-density layer in reflectivity data. The porosity of this layer decreased with the increase in the film thickness. In the thicker films, 67.5 and 97 nm thick, no porous layer is observed the scattering length density variation at the surface is smooth and no fluctuations are observed. The present study clearly demonstrates that morphology of ultra thin films of NiMnSb is markedly different from the films with thickness greater than 60 nm as seen by scattering length density variation at vacuum film interface. The distinctly different magnetic and transport properties observed for thickest (D) and thinnest (A) samples show that this morphological disorder influences the physical properties of these films. The XRF studies rule out any compositional anomaly and the influence of surface/interface roughness as the source of the interesting magnetic and transport properties. It will now be interesting to investigate the electronic structure of these off-stoichiometric NiMnSb films.

\section{Acknowlwdgements}

The authors from ISEL/FORTH would like to acknowledge support from the EU project FENIKS G5RD-CT-2001-00535.

{99}                        

\end{document}